\documentclass[journal, 12pt, twocolumn, final]{IEEEtran}
\usepackage{algorithmic}

\usepackage{latexsym}
\usepackage{epsfig}
\usepackage{amsfonts,amsmath,amssymb,amstext}
\usepackage{epstopdf}
\usepackage{graphicx}
\usepackage{psfrag}
\usepackage{dsfont}
\usepackage{color}
\usepackage{soul}
\usepackage{cite}
\usepackage{mathrsfs}
\usepackage{subcaption}
\usepackage[singlespacing]{setspace}

\begin{document}
\title{Reconfigurable Intelligent Surfaces\\ in Challenging Environments:\\ Underwater, Underground, Industrial and Disaster}
\author{Steven Kisseleff, ~\IEEEmembership{Member, IEEE}, Symeon Chatzinotas,~\IEEEmembership{Senior Member, IEEE}\\ and Björn Ottersten,~\IEEEmembership{Fellow, IEEE}
\thanks{This work was supported by Luxembourg National Research Fund (FNR) under the CORE project RISOTTI C20/IS/14773976.\newline
The authors are with Interdisciplinary Centre for Security, Reliability and Trust (SnT), University of Luxembourg, 1855 Luxembourg. Corresponding author: Steven Kisseleff (e-mail: steven.kisseleff@uni.lu).}
}

\maketitle
\thispagestyle{empty}
\begin{abstract}
Challenging environments comprise a range of scenarios, which share the fact that it is extremely difficult to establish a communication link using conventional technology due to many impairments typically associated with the propagation medium and increased signal scattering. Specifically, underwater and underground media are known to absorb electromagnetic radiation, which heavily affects the overall path loss. Industrial and disaster environments can be viewed as rich scattering environments with corresponding substantial multipath propagation leading to intersymbol interference and deterioration of signal quality. Although the challenges for the design of communication networks, and specifically the Internet of Things (IoT), in such environments are known, there is no common enabler or solution for all these applications.

Reconfigurable intelligent surfaces (RISs) have been introduced to improve the signal propagation characteristics by focusing the signal power in the preferred direction, thus making the communication environment 'smart’. While the usual application of RIS is related to blockage avoidance, the very same technique can be used to reduce the effect of multipath and even partially compensate the signal absorption via passive beamforming.

Due to the beneficial properties of RIS, its use in challenging environments can become the aforementioned enabler and a game changing technology. However, various aspects of RIS deployment and system design need to be addressed in order to fully benefit from this technology. In this paper, we discuss potential use cases, deployment strategies and design aspects for RIS devices in underwater IoT, underground IoT as well as Industry 4.0 and emergency networks. Furthermore, we provide a potential hardware architecture and derive the expected signal quality improvements with increasing number of RIS elements. The numerical results reveal substantial performance gains of up to 20 dB per decade. In addition, novel research challenges to be addressed in this context are described.
\end{abstract}


\section{Introduction}
\label{sec:1}
\subsection{Motivation}
\IEEEPARstart{R}{econfigurable} intelligent surfaces (RISs) is a promising technology, which has been motivated by the increasing demands of wireless networks beyond 5G~\cite{Liaskos}. RISs are intelligently designed artificial planar structures with reconfigurable properties enabled via integrated electronic circuits. These electronic circuits can be programmed to reflect an impinging electromagnetic (EM) wave in a controlled manner, thus making the communication environment controllable and smart. RISs are manufactured with low-profile and light weight using inexpensive materials and can be deployed on a variety of surfaces, such as facades of buildings, walls, ceilings, etc.~\cite{Coquet2018}. The signal propagation from transmitters to receivers can therefore be assisted by steering the RIS-reflected signals in directions that enhance the signal quality, suppress co-channel interference and consequently improve the spectral efficiency. The corresponding performance gains of RIS-assisted networks compared to current wireless systems are well-known from the previous works in this domain~\cite{Renzo2019}. Furthermore, co-channel interference can be reduced or avoided by choosing different propagation paths for the interfering signals. This is important in urban environments with dense deployment of terminals, cf. \cite{kisseleff2020reconfigurable}.

Challenging environment (CE) refers to a communication environment, in which the signal propagation and connectivity using traditional signaling technologies are heavily impaired, so that a stable, reliable and secure communication link is very difficult to establish. Unlike in the so-called extreme environments, the challenges in CEs are mostly related to wireless communication rather than to the operability of sensor nodes under extreme conditions, cf. \cite{rashvand2017wireless}. Depending on the actual environment, the reasons for the impaired connectivity can be very different. One of the typical problems is the increased signal scattering and absorption. Through this, the path loss increases with transmission distance much faster compared to traditional communication environments, i.e. path loss exponent higher than 4. Another problem is related to the accessibility of the network nodes and security of the links. In this context, it is important to find either the shortest or the least impaired path to the intended receiver. With the introduction of the Internet of Things (IoT) in CEs, the problem of establishing a reliable communication link between the IoT nodes has become a real challenge, which has no unified solution yet.

Among various challenging and extreme communication environments, in this work, we focus on four use cases: underwater, underground, industrial, and disaster environments. These environments have very different characteristics and challenges. These problems of the CEs typically motivated the researchers to look for alternative communication-enabling technologies which are very different as well. As an example, magnetic induction has been proposed for the signaling in the underground medium, since it is less prone to scattering and absorption than the traditional EM waves. For the same reason, acoustic signaling has been selected for the underwater communication. Correspondingly, each use case has evolved into an individual research field with specific system design methods and limited generalization potential. Moreover, the interconnection of various IoT networks, such as disaster and industrial environments or underground and underwater, is hardly possible due to the divergent design guidelines and techniques, which prohibit joint optimization. Similarly, it is very challenging to incorporate these networks into terrestrial networks and to achieve homogeneity in their design.

To solve this issue, RIS can be proposed in nearly all possible CEs in order to improve the connectivity. Using the additional degrees of freedom associated with RISs, it is therefore possible to determine the optimal signal path and reduce the impact of the environment on signal propagation. Hence, RIS has the potential of becoming the key enabler for the future wireless systems in CEs. It might be even possible to unify the design methods for the wireless systems in various CEs based on this common enabler. This might be possible via conversion of the individual challenges into common challenges for all environments. In this context, future research on RIS-assisted CEs becomes very promising and interesting. 

In this work, we aim at providing our vision on the application of RIS in CEs in order to open the discussion on the evolution of RIS-assisted CEs and pave a way to their systematic design.
\subsection{Contributions}
Throughout the open literature, the research challenges of integrating RISs into wireless networks are solely discussed within the existing network architectures and the traditional urban ecology. These challenges differ from the ones discussed in the context of CEs, which motivates us to fill this gap by providing a vision of RIS-assisted CE networks along with the proposal of some relevant research challenges. Specifically, we discuss relevant application scenarios, deployment strategies and aspects of the system design. The contributions of this work provide a basis for the future investigations in this field as it points in the direction of the most relevant research problems to be addressed in the near future.
The main contributions are as follows:
\begin{itemize}
    \item A brief review of the RIS technology is provided including the application of RISs with unconventional signaling techniques, such as magnetic induction and acoustic waves;
    \item An overview of the peculiarities and features of each considered environment is provided in order to identify the main challenges for the wireless networking without RIS. Specifically, the frequency-selective signal absorption in underground and underwater environments as well as heavy multipath and signal blockage in industrial and disaster environments are considered;
    \item Potential benefits from the deployment of RIS in the respective environment are explained, i.e. reduced frequency selectivity, improved coverage, reliability and energy efficiency of the communication network;
    \item Deployment strategies for RIS as part of respective IoT networks are discussed. In this context, novel suitable configurations of RIS for omnidirectional operation in underwater environment and for separable operation in disaster environment are proposed;
    \item Generic hardware design is proposed to account for the autonomous operation of RIS and possible energy and processing cooperation among RIS devices;
    \item Expected improvements of the signal quality are derived for the generalized CE. Furthermore, for each considered CE, we select a practical scenario and evaluate the signal quality with and without RIS, thus confirming the analytically derived performance gains;
    \item Future research challenges for the unified RIS-based technology in CEs are revealed.
\end{itemize}
This paper is organized as follows. Prior works and recent advances of the RIS technology are discussed in Section \ref{sec:2_new}. Sections \ref{sec:underwater}-\ref{sec:disaster} provide overviews and describe the peculiarities for the application of RIS in the underwater, underground, industrial and disaster environments, respectively. Specifically, the most beneficial strategies for the RIS deployment in these scenarios and the respective challenges for the signal propagation and system design are explained. In Section \ref{sec:system}, a promising hardware concept for the RIS application in CEs and a generic system model is proposed. Furthermore, the expected signal quality in RIS-assisted communication in CEs is derived and numerically evaluated. Future challenges and research opportunities are outlined in Section \ref{sec:4}. Finally, the paper is concluded in Section \ref{sec:5}.
\section{Advances of RIS technology}
\label{sec:2_new}
Communication channels have been traditionally considered as an uncontrallable entity. Hence, most of the conventional wireless communication techniques attempt to adapt the system parameters to the peculiarities of the uncontrolled channel. However, substantial performance gains can be achieved, if the communication channel becomes reconfigurable and correlates with the signaling. This idea has been realized using RIS, which can be used to enhance the performance of communication systems by controlling the reflections of the impinging signals.


RIS has been named a new paradigm in wireless communications in~\cite{Liaskos}. The authors proposed RIS as one of the key enablers for the 5G and beyond systems. In particular, research challenges and directions for the design and deployment of RIS based on the prior work~\cite{Yang} have been discussed. Furthermore, the idea of controllable communication environment has been proposed in various works under different names, such as Intelligent Reflective Surface (IRS), Large Intelligent Surface (LIS), Intelligent Mirrors, Intelligent Metasurfaces ~\cite{Hu, Huang, Gong}.

\subsection{Promising research directions}
\label{sec:sota}
Since RIS represents a fundamentally new concept of signal propagation, channel modeling and acquisition are of utmost importance for the system design. Hence, multiple works have been dedicated to these topics, e.g.~\cite{Zheng2019,jensen2019,zheng2020uplink, wang2020channel,wei2020channel,elbir2020deep}. Since RIS can actively modify the communication environment, the proposed channel estimation methods depend on the individual scenarios, number of RISs, users, etc. \cite{wei2020channel,zheng2020uplink}. 

Furthermore, advanced techniques of multi-antenna transmissions need to be developed. Specifically, novel precoding and beamforming methods have been proposed for RIS-assisted multiuser networks \cite{wu2019intelligent,zhou2020robust,zhang2020joint,liu2019joint,liu2020ris,yan2020passive,li2020reconfigurable}. In this context, joint passive and active beamforming have been investigated for various system configurations. It has been shown that optimized reflective coefficients provide substantial performance gains in terms of spatial diversity and quality of service. Statistical improvements of the spatial diversity and especially the matrix rank have been addressed in \cite{delHougne} and validated in practical experiments.

Most of the relevant works in the research field of RIS focus on indoor scenarios, where the reflections from the walls can be potentially very harmful without RIS. Using RIS, the number of connected devices in such scenarios can be increased without causing a performance degradation for the network due to harmful co-channel interference signals. Note that the indoor deployment of RIS provides the biggest advantage and diversity gain due to relatively short transmission distances, such that the attenuation of the reflection-based signal paths is not significantly higher than the direct path. 

Outdoors, the main target application will be Smart Cities, cf. \cite{kisseleff2020reconfigurable}. So far, however, not many works have investigated RIS-assisted outdoor scenarios. The main innovation addressed in these works is the use of unmanned aerial vehicles (UAVs)~\cite{Ma, Li,yang2020performance,lu2020aerial}. In such scenarios, UAVs can act not only as receivers, but also as mobile relays to serve as access points for multiple network nodes. As an example, the beamforming towards UAVs and the tracking of the UAV position by the BS has been investigated in~\cite{Li}.

High frequency bands have been viewed as especially attractive for the RIS-assisted wireless networks. Specifically,  Millimeter-Wave (mmWave), Terahertz (THz) and optical signaling are characterized by a very small wavelength, which leads to rather small transmit antennas and increased reflection of the impinging waves from the surface compared to the traditional carrier frequencies of a few GHz. Accordingly, many works focus on THz and mmWave based technologies, cf. \cite{xiu2020irs,wang2020intelligent}. However, recent studies have been conducted to introduce RIS in optical communication networks, cf. \cite{najafi2019intelligent}. Here, the weak absorption of light can be exploited in order to guide the signal to the destination with a very little path loss.

An additional motivation for the theoretical investigations in this area has been provided in~\cite{welkie2017programmable,Cui,delHougnePhysRevX}, where the first experimental results for RIS-assisted signal propagation have been presented. Recently, more thorough practical design aspects and experimental studies have been published, cf. \cite{dai2020reconfigurable,9133157,tang2020wireless,9133266,9053976}. Specifically, in \cite{dai2020reconfigurable}, detailed guidelines for the hardware design of RIS as well as the complete signal flow model has been presented. In \cite{9133266}, a hardware design for RIS configuration as transmitter/modulator has been proposed and implemented. However, the main drawback of these existing designs lies in the assumed accessibility of the RIS devices, which is not true in CEs. Hence, more advanced functionalities, such as autonomous operation and energy independence need to be incorporated into the hardware design. In Section \ref{sec:hardware}, we propose such an architecture for RIS in CEs.

Promising new research directions, which have been addressed recently comprise:
\begin{itemize}    
    \item path loss modeling \cite{tang2020wireless,danufane2020path,basar2020simris}
    \item physical layer security \cite{chen2019intelligent,du2020reconfigurable,ai2020secure},
    \item non-orthogonal multiple access (NOMA) \cite{hou2020reconfigurable,elhattab2020reconfigurable,li2020joint},
    \item wireless power transfer and simultaneous wireless information and power transmission (SWIPT)\cite{yang2020reconfigurable, pan2019intelligent, zhao2020wireless, wu2019weighted},
    \item vehicular communication \cite{kisseleff2020reconfigurable, ai2020secure,huang2020transforming,masini2020use,ozcan2020reconfigurable},
    \item cell-free networking \cite{zhang2020joint,zhang2020beyond,huang2020decentralized},
    \item localization \cite{abu2020near,wymeersch2020radio,elzanaty2020reconfigurable}
    \item RF sensing \cite{9133157,9141218}.
\end{itemize}
The research in these directions is very active, such that multiple high-quality publications appear on a daily basis. For more insight in the advances in RIS technology and the creation of a smart environment, we refer the reader to~\cite{Renzo2019,Gong,di2020smart,liu2020reconfigurable,pan2021reconfigurable}, where all relevant studies have been described and future challenges in general context have been addressed.


\subsection{Alternative signaling technologies}
In this work, we consider the most CEs for the wireless communication. In some of these CEs, such as industrial and disaster environments, the traditional EM waves in the moderate and high frequency bands can be directly employed. In others, such as underwater and underground environments, EM waves may not provide a sufficient signal quality. In these environments, alternative signaling technologies have been proposed to overcome the drawbacks of the EM waves, such as absorption and scattering. Hence, it is worth looking into the advances of RIS specifically designed with respect to these technologies in order to understand their benefits and practical limitations.
\begin{figure*}[t!]
    \centering    
    \begin{subfigure}[t]{0.4\textwidth}
    \includegraphics[width=0.9\textwidth, height=0.25\textwidth]{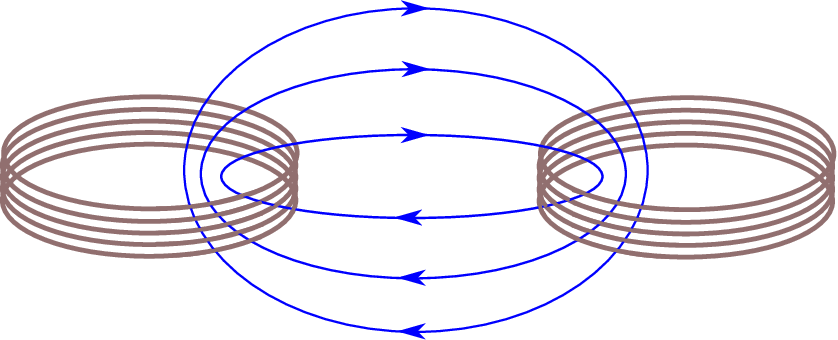}
    \caption{Transceivers for magnetic signaling.}
    \label{fig:magnetic}
    \end{subfigure}
    ~
    \begin{subfigure}[t]{0.4\textwidth}
    \includegraphics[width=0.9\textwidth, height=0.25\textwidth]{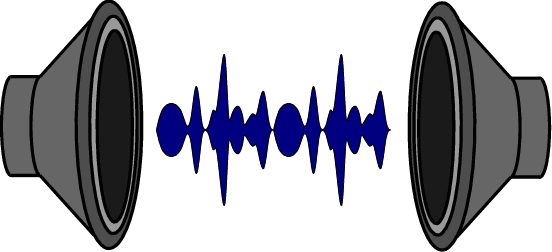}
    \caption{Transceivers for acoustic signaling.}
    \label{fig:acoustic}
    \end{subfigure}
    \caption{Transceivers for alternative signaling technologies.}
\end{figure*}
\subsubsection{Magnetic induction/resonance}
Magnetic induction (MI) or magnetic resonance refers to non-propagating quasi-static magnetic fields using induction coils as part of respective resonance circuit instead of traditional EM antennas, see Fig. \ref{fig:magnetic}. Typically, MI is used with very low signal frequencies, i.e. between few kHz (in conductive media) and 13.56 MHz (according to Near-Field Communications standard), which results in very narrow frequency band. Due to the low frequency, the signal wavelength is very large and does not suffer from reflections or scattering. Unfortunately, this fact prevents from directly utilizing the RIS technology with MI. Nevertheless, a similar effect can be achieved via passive relaying of magnetic fields using the so-called MI waveguides, cf. \cite{shamonina2002magneto}. Specifically, passive MI coils aligned in a waveguide structure generate secondary magnetic fields without explicit power consumption, which can be viewed as pseudo-reflections. These pseudo-reflections can be made reconfigurable by modifying the resonance circuit impedance of each MI relay \cite{6567880}. Hence, these passive MI relays can be viewed as a special case of magnetic RIS with a single reflective element per device. Note that even with a single element RIS can typically contribute to the signal propagation in a network by manipulating the angle of reflection leading to changes of the phase and magnitude of the reflected signal. In case of MI relay, the pseudo-reflection can be similarly adjusted to change the phase and the magnitude of the reflected signal. 
\subsubsection{Acoustic communications}
Depending on the application, acoustic signals are typically generated in the frequency range between few Hz (infrasound) and few MHz (ultrasound). Due to the different type of signal carrier (pressure rather than EM fields), the signals are transmitted using speakers and received using a microphone (or hydrophone in the water medium), see Fig. \ref{fig:acoustic}. Although the conventional RIS technology has been proposed for EM waves, the possibility of redirecting acoustic waves has been introduced in \cite{zhao2013redirection}. Recently acoustic RISs have been proposed, cf. \cite{fan2019tunable,fan2020reconfigurable,li2020arbitrarily}. The primary scope of these works was the cloaking of uneven surfaces to avoid the scattering and improve the ambient sound. However, the use of such acoustic RIS devices for communications has been recognized in \cite{tian2019programmable}, where the possibility of acoustic beamforming using specifically designed metasurfaces has been explored. Furthermore, the application of RISs for the underwater communication has been demonstrated in \cite{he2020experimental} using experimental setup. Unlike EM and MI based wireless systems, the passive signal steering or beamforming are not achieved via circuit impedance reconfiguration, but via actual mechanical manipulation of the surface elements \cite{fan2020reconfigurable}.

\subsection{Challenging environments}
In the following sections, we discuss four most relevant CEs. The importance of some of these environments and their characteristics have been discussed in \cite{rashvand2017wireless,8543577}. We start with the underwater and underground communications, where the propagation medium is the main challenge for the system design. Then, we move on to the industrial and disaster environments, where the large number of obstacles impacts the signal propagation. Note that we do not address higher-layer challenges for the network design such as routing and topology control, which depend on the actual applications and system configuration. Instead, we focus on the challenges associated with the connectivity and digital signal transmission. Nevertheless, high OSI layers will be certainly addressed in future research on RIS-assisted communication both in traditional environments and in CE.
\section{Internet of Underwater Things (IoUwT)}
\label{sec:underwater}
\subsection{Overview}
Underwater communication networks have been analyzed in a plethora of works for various applications. The research in this area can be traced back to the early days of wireless communications, cf. \cite{moore1967radio}. 

The main difference between underwater networks and the terrestrial networks lies in the propagation medium, which dramatically affects the path loss. Among the main relevant properties of the water medium, its conductivity is by far the most important one. The conductivity of water depends on its salinity, temperature and pressure, such that it has values of around 0.01 S/m in sweet water and up to 4-5 S/m in sea water, cf. \cite{popovic2000introductory,8763947}. These high conductivity levels lead to absorption of the EM fields and very high path loss. Thus, data rates above 8 kbit/s are prohibited over distances beyond 10 m in the sea water \cite{che2010re}.

For this reason, underwater communications have been traditionally designed using acoustic waves, which do not suffer from absorption as much as the other signaling techniques. Acoustic signaling relies on the propagation of pressure variations, which can reach up to tens of kilometers in the open water. However, the issues with the acoustic signaling are related to the typical maritime objects, such as ground/shore or fish, to the interference from dolphins/whales, and to the signal scattering/reflections in the acoustically heterogeneous medium, i.e. water streams with different pressure and temperature levels. The latter is due to the varying speed of sound, which depends on the local properties of the medium according to \cite{8763947}: $c=1412+3.21T+1.19S+0.0167z$, where $S$ is the salinity, $T$ is the temperature and $z$ is depth. In addition, the absorption loss is substantial. All these issues impact the communication channels by means of multipath propagation, frequency-selective and time-variant impulse responses as well as interference. Alternative solutions based on extremely low signal frequencies using MI or based on optical communication have been introduced as well \cite{li2019survey,kaushal2016underwater}. However, these technologies are much less common due to the path loss, which is not necessarily lower than using traditional EM waves, cf. \cite{che2010re}.

\begin{figure}
    \centering
    \includegraphics[width=0.48\textwidth]{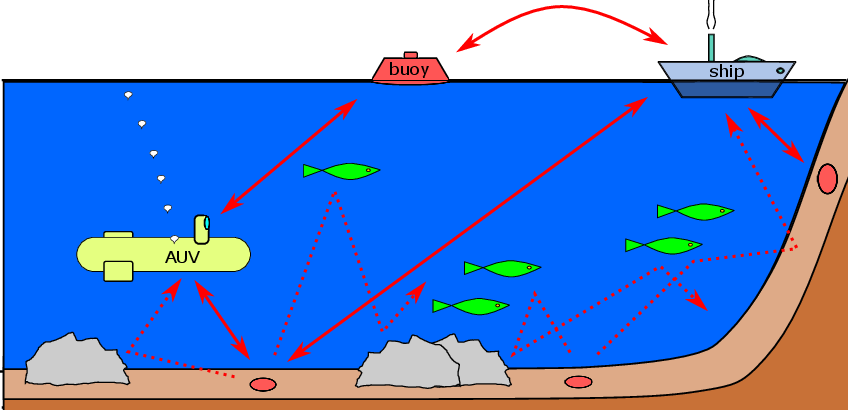}
    \caption{Example of the IoUwT. Connectivity between networks nodes, i.e. submarine, buoy, ship and sensors, is impaired by reflections from the typical maritime obstacles like fish and rocks.}
    \label{fig:underwater}
\end{figure}

Recently, the idea of establishing a large wireless network similar to the upcoming IoT has been introduced. It has been named the Internet of Underwater Things\footnote{The usual abbreviation of the Internet of Underwater Things is IoUT, which is commonly used in the literature. However, in this work, we consider also the Internet of Underground Things, which would have the same abbreviation. Hence, in order to distinguish between the two types of network, we utilize IoUwT for the underwater networks and IoUgT for the underground networks.} (IoUwT) and should connect various devices deployed under the water surface, e.g. submarines, autonomous underwater vehicles (AUVs), sensors and ships, see Fig. \ref{fig:underwater}.
Here, the goal is to provide sufficient connectivity for the wireless network in the underwater medium. Usual scenarios for the deployment of the IoUwT involve the observation of maritime animals and fish, monitoring of seismic activities (e.g. tsunami) and wireless communication with underwater vehicles. Sometimes, the network includes floating buoys to collect the data from the sensors deployed below the surface. In all these scenarios, the mentioned issues of time-variant multipath channels (when using acoustic waves) seem to be the main challenge for the communication.

IoUwT is a promising research area, which has seen a lot of advances in the recent years. For more information on underwater communications and IoUwT we refer to \cite{domingo2012overview,8763947}.

\subsection{RIS in Underwater medium}
While EM, MI and even optical signal propagation in the water medium (especially in clean sweet water) are possible over short distances, transmissions over larger distances using acoustic signaling are preferred in the context of the IoUwT. With large distances, the absorption of EM radiation by the medium leads to a heavy degradation of the signal quality, which cannot be combated using RIS technology. Hence, we focus on acoustic signaling in this application. 

\begin{figure}
    \centering
    \includegraphics[width=0.48\textwidth]{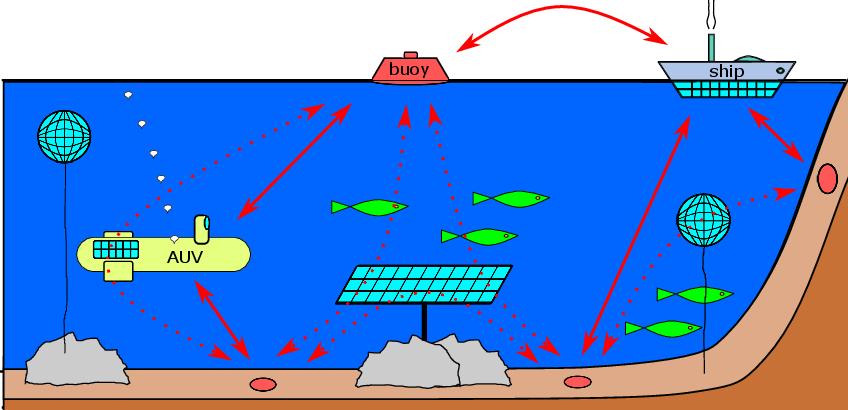}
    \caption{IoUwT assisted by RIS.}
    \label{fig:underwater_RIS}
\end{figure}

As mentioned earlier, acoustic signal propagation in the underwater medium suffers most from scattering at uneven surfaces, water streams and fish. This leads to a high path loss and extreme frequency selectivity of the communication channel due to the multipath, such that only very narrow signal bandwidths are tolerated. Correspondingly, the effective data rate is very low. Hence, these weaknesses need to be mitigated using RIS. Note that the role of RIS in this scenario is not to avoid blockage by finding a more suitable path around it (as it is usually suggested in conventional terrestrial communications), but to combat the multipath effect, which is present even in case of line-of-sight propagation due to frequency-selective fading in the water medium. The deployment of RIS in this scenario is illustrated in Fig. \ref{fig:underwater_RIS}. 
\subsubsection{General approach}
The steering of the signals in the preferred direction would help constructively align the interference and reduce the multipath. Of course, it may not be possible to completely remove the multipath due to a large number of possible reflections in the water. However, the orientation of RIS elements can be optimized to reduce the frequency selectivity and thus increase the effective signal bandwidth. In this context, the design problem can be formulated as a maximization of the signal quality measured after the applied practical equalization filter in the receiver. Since the signal quality is typically related to the discrete length of the equivalent impulse response (which is associated with the maximum delay spread according to the multipath), an alternative objective can be formulated as a minimization of the delay spread. The solution to this optimization problem needs to be updated very quickly due to the typically short coherence time in the water medium. Hence, the corresponding computational complexity needs to be very low in this scenario. Thus, learning-based solutions, e.g. deep learning, seem especially promising, since the most computationally expensive part of the optimization is performed offline. Alternatively, robust optimization can be applied in order to avoid the necessity of rapid updating the phase shifts.

The deployment of RIS in the underwater medium can have various configurations depending on its location:
\begin{itemize}
    \item attached to the ground or to the shore;
    \item attached to autonomous underwater vehicles;
    \item floating beneath the surface.
\end{itemize}

Next, we would like to discuss some of the challenges and potential solutions related to the operation of the RISs in these three underwater scenarios.
\subsubsection{Stationary deployment}
The first option, i.e. stationary deployment of RIS, is the simplest and the most beneficial option, since it does not pose too many new challenges for the system design. Hence, RIS can be readily integrated into the IoUwT.

The main benefit of the stationary deployment is that the geometrical configuration does not change over time. In principle, this should lead to more or less stationary communication channels, which is however only valid for relatively short distances. For large distances, e.g. above 100 m, due to the fast time-varying channels in the underwater medium, it might be difficult to reach a sufficiently high level of synchronization and channel estimation accuracy in order to fully benefit from the capabilities of RIS. Hence, robust optimization of the orientation of RIS elements seems to be the key solution that helps to mitigate the impact of the time-varying multipath channels (to some degree) at least on average. 

\subsubsection{Autonomous underwater vehicles (AUVs)}
AUVs are expected to serve the deployed nodes of the IoUwT by moving from one node to another, which resembles the traditional mobile relaying. While the main task is to ensure a better connectivity and to recharge the batteries of the distant nodes, AUVs are well suited for carrying RIS. Through this, they can contribute to the smart underwater environment. This type of mobility is therefore steerable and can be adjusted to improve the signal propagation using phase shifting capabilities of RIS in addition to the mobility of the vehicles. In this context, the trajectory of the AUVs is of special interest.

One of the challenges related to the mobility scenarios is the channel prediction. Unlike with stationary deployment, the geometry of the system configuration is not fixed, but changing. Correspondingly, the communication channels may become time-varying. However, the motion of the AUVs may be relatively slow, which leads to a rather long motion-related coherence time. Hence, the motion is not crucial for the system assisted by RIS-equipped AUVs.

On the other hand, the surface of the AUVs will be probably designed to reduce the friction in water, which may otherwise become the main problem for the steering of the vehicle. Hence, only relatively small areas should be covered by RISs in order to reduce this harmful effect while preserving the control over the smart environment.

\subsubsection{Floating RIS}
One of the most promising deployment strategies involves a floating RIS. In this scenario, RIS can be attached to the ground by a cable, see Fig. \ref{fig:underwater_RIS}. On the one hand, this allows RIS to be placed at the desired depth, e.g. between the sensors deployed on the ocean ground and the surface, and on the other hand restricts the maximum distance between the anchor position and RIS. Also, the cable connection can be used for the control signaling, which would substantially simplify the design of RIS, since the control signal does not need to be transmitted wirelessly.

\begin{figure}[t!]
    \centering
    \begin{subfigure}[t]{0.2\textwidth}
    \includegraphics[width=\textwidth,height=\textwidth]{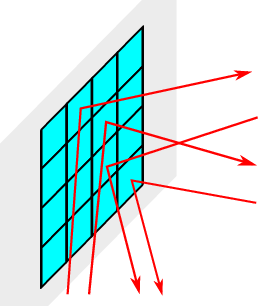}
    \caption{Planar RIS.}
    \label{fig:planar}
    \end{subfigure}
    ~
    \begin{subfigure}[t]{0.2\textwidth}
    \includegraphics[width=0.9\textwidth, height=0.9\textwidth]{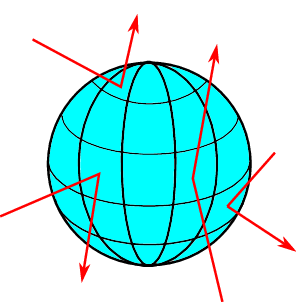}
    \caption{Spherical RIS.}
    \label{fig:round}
    \end{subfigure}
    \caption{Possible designs of RIS for underwater scenario.}
\end{figure}

Unlike traditional RIS deployment scenarios, i.e. smart buildings, smart homes, etc., where the impinging signals arrive always from the same half-space, see Fig. \ref{fig:planar}, the floating RIS should be able to reflect signals coming from all directions. 
Hence, for this specific type of RIS, a novel design is required. In particular, the deployment of reflective RIS elements on the surface of a sphere seems promising, see Fig. \ref{fig:round}. But other geometrical bodies can be considered as well depending on the specific application. Similarly to the planar RIS, each element can reflect the impinging signals in a controlled manner. Nevertheless, the number of reflective elements, which can be used to steer the signals in the same direction is reduced compared to planar RIS, such that the benefit from the spherical RIS is in general lower than with planar RIS. In order to avoid the performance degradation, the size of RIS needs to be substantially increased compared to the planar RIS. Through this, a similar number of elements pointing in similar direction can be accommodated, which implies a similar performance. However, this drawback is compensated by the very advantageous property of the spherical RIS, i.e. the operation of RIS independently of its orientation. This property is very important for the application in underwater scenarios due to potential rotational motion of RIS in water streams or after collisions with other objects.
\section{Internet of Underground Things (IoUgT)}
\label{sec:underground}
\subsection{Overview}
The research on communications in the underground medium is nearly as old as that of the underwater communications. Similarly, the main challenge of this environment is related to conductive propagation medium, although the conductivity is typically much lower compared to the water medium, below 0.01 S/m. The soil conductivity and correspondingly the absorption rate for the EM radiation depend on the so-called volumetric water content (VWC), i.e. the relative amount of water in soil, as well as on the soil content, such as sand, clay, etc. \cite{vuran2010communication,7471430}. Furthermore, variations of the soil content can lead to signal reflections and scattering. These effects lead to a severe multipath and very high path loss. Besides, signal reflection at the ground surface due to the abrupt change of the EM properties contributes to the multipath propagation as explained in \cite{vuran2010communication}.

Due to much lower average soil conductivity, most of the works on underground communications are based on traditional EM waves despite the high path loss. However, it is still extremely difficult to reach transmission distances above 50 m. Moreover, low frequency MI solutions have been proposed to overcome the issues related to the low coverage, cf. \cite{MI_comms_WUSN, 8464253}. The drawback of the latter is however due to a very narrow bandwidth (centered around the resonance frequency) and relatively large transceivers. 

\begin{figure}
    \centering
    \includegraphics[width=0.48\textwidth]{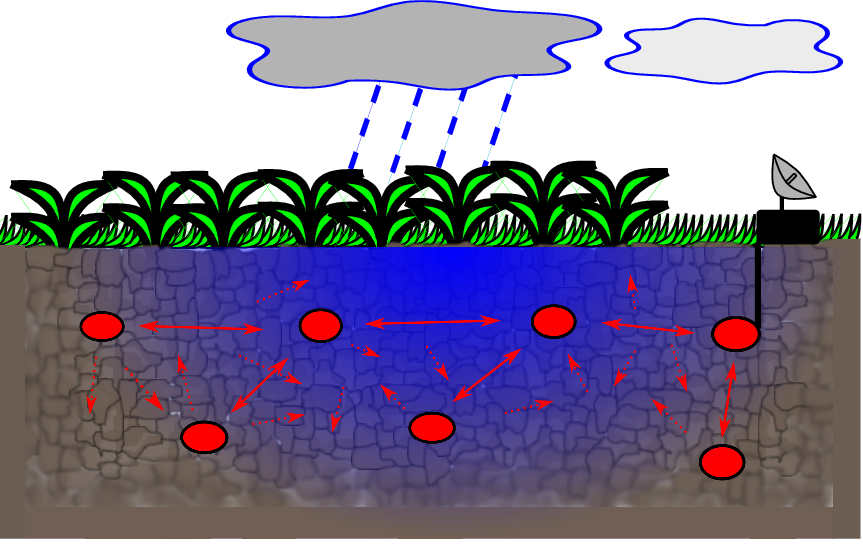}
    \caption{Example of the IoUgT for smart agriculture applications. The sensors are deployed below the ground surface and monitor the water content in the soil. The signals are scattered by the heterogeneous soil medium.}
    \label{fig:underground}
\end{figure}

In the context of both EM waves and MI, there has been much progress in the design of wireless networks lately. In particular, wireless underground sensor networks (WUSNs) have been proposed \cite{akyildiz2006wireless}, which evolved into a concept of the Internet of Underground Things (IoUgT) in the recent years \cite{saeed2019toward}. Here, the goal is to establish wireless connectivity between distributed sensor nodes in order to monitor seismic activity, structural health of buildings as well as soil quality for the smart agriculture and clever irrigation, see Fig. \ref{fig:underground}. 
The main challenge for this type of wireless networks remains the medium-dependent path loss, which leads to a low coverage area and low signal quality.

Furthermore, such wireless networks can be used to provide a communication infrastructure to hardly accessible areas like mines and tunnels. In such scenarios, signal reflections contribute on the one hand to the signal propagation, since each tunnel can be viewed as a waveguide, and on the other hand to the amount of intersymbol interference, which drastically increases with each reflection. This can lead to a poor signal quality at the receiver.

Similarly to the IoUwT, the research on IoUgT is very promising and has been popularized in the recent years. For more information on IoUgT, we refer to \cite{saeed2019toward,salam_book}.

\subsection{RIS in Underground medium}
For the underground communications, we focus on signaling using traditional EM waves in the moderate frequency range. While MI based underground communication represents a promising alternative, the potential of magnetic RIS has been thoroughly studied and did not show sufficient performance improvement to justify the deployment of additional resonant circuits under realistic assumptions \cite{7093165}. The reason for this seems to be the limitation of the number of reflective elements which can be used in parallel by one magnetic RIS. Also, acoustic signaling is not possible due to severe path loss and scattering effects inside the acoustically heterogeneous soil medium.

The signal propagation in the underground is very vulnerable to the heterogeneous conductive medium of soil and in particular to the VWC as pointed out earlier. Additional effects like scattering and reflections are related to the soil composition. Similarly to the underwater communication, these effects lead to a very high EM path loss even at short transmission distances, cf. \cite{vuran2010channel}. Furthermore, multipath propagation results directly from the signal reflections and can be very harmful for the receive signal quality. Hence, the main challenge that motivates the application of RIS in this context is to reduce the multipath and the path loss by avoiding the signal scattering and by steering the signal through the soil layers with low VWC. Similar to the underwater scenario, this challenge exists even in case of potential line-of-sight propagation.
\subsubsection{General approach}
\begin{figure}
    \centering
    \includegraphics[width=0.48\textwidth]{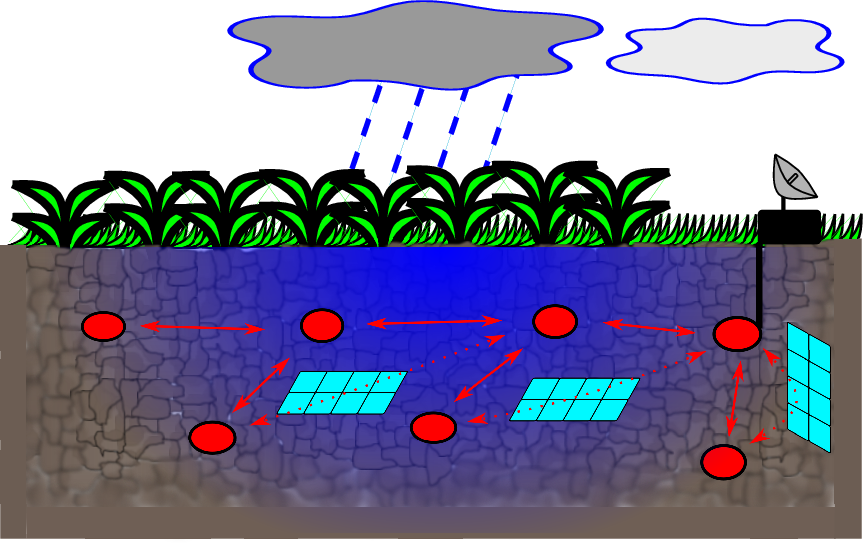}
    \caption{IoUgT assisted by RIS.}
    \label{fig:underground_RIS}
\end{figure}

An example for the deployment of RIS in the underground environment is illustrated in Fig. \ref{fig:underground_RIS}. 
Similarly to the application of RIS in the underwater medium, the phase shifters of RIS can be optimized to reduce the frequency selectivity and increase the effective signal bandwidth. Correspondingly, a practical design problem would aim at minimizing the delay spread and signal quality after the equalization filter. Furthermore, the general challenges associated with RIS remain mostly the same as with underwater communications. Correspondingly, the solution based on machine learning seems promising for this use case, too. However, the main difference lies in the substantially longer coherence time in the underground medium. While the time-varying channels in the underwater medium are much more dynamic due to the water streams, fish, etc., the underground medium can remain constant for a very long period of time, e.g. between two rain periods. Even during the rainy weather, the VWC in soil does not change abruptly, but requires several minutes to provide a measurable change in the signal propagation characteristics. Hence, the complexity of the optimization problems does not need to be as low as in the underwater medium. Specifically, the parameter tuning may need to be executed only a few times in the entire lifetime of the designed IoUgT, since the transmission channels remain mostly unchanged, such that the solution remains unchanged as well. Through this, a more thorough channel estimation and more exact solution can be obtained using traditional convex optimization techniques.
\subsubsection{Deployment in tunnels and mines}
\begin{figure*}
    \centering
    \includegraphics[width=0.8\textwidth]{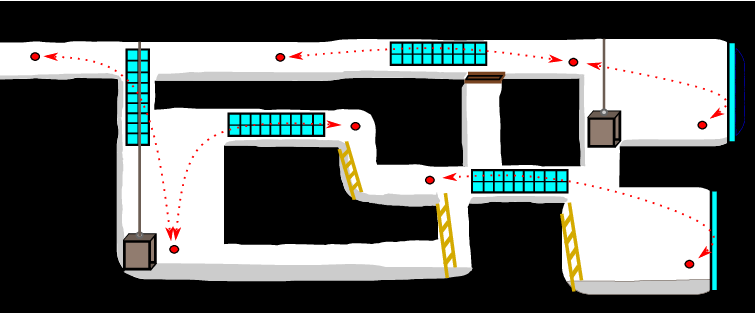}
    \caption{Mines and tunnels scenario. RISs are deployed in the walls to improve the directivity of signal propagation.}
    \label{fig:tunnel}
\end{figure*}

One of the main application scenarios for the IoUgT is the deployment of sensor nodes in mines and tunnels. The main challenge for the EM signal propagation is related to the uncoordinated reflections from the walls of the tunnel or mine as well as scattering at the edges of the walls. This leads to the performance degradation for the signal quality and connectivity of the network. To address this issue, RISs can be installed in the walls and the ceiling of the tunnel in order to improve the directivity of signal propagation by steering the signal in the passage direction, see Fig. \ref{fig:tunnel}.

Furthermore, as mentioned earlier, soil may partially absorb the signal, such that each reflection potentially contributes to the overall path loss. On the other hand, RISs are typically designed with the maximized reflection coefficient in order to avoid additional losses, which would otherwise render the use of RIS inefficient. Hence, the main challenge for the design of RIS-assisted communication in tunnels and mines is to determine the optimal strategic placement of the RISs and their phase shifts, which would guide the signal in a ping-pong manner through the passage. Due to the stationary deployment of the IoUgT, the optimal position and coarse values of the phase shifts can be determined very accurately offline using solely the map of the passage. 

\subsubsection{Other scenarios}
In other scenarios of the underground communication networks, such as soil monitoring, the nodes are typically embedded in the soil medium, such that the communication between them is established directly through the ground. In this context, the use of large RISs may not always be feasible due to a high deployment effort. On the other hand, the effectiveness of RIS increases with the number of RIS elements, which typically scales with the RIS size, since it enables a higher spatial diversity \cite{delHougne}. Hence, it is important to investigate a potential tradeoff between the performance gain and the deployment costs with respect to the number of RIS elements and deployment strategies. In some cases, a large set of small RISs may be preferable compared to a single large RIS, leading to a distributed RIS operation\footnote{This problem occurs in disaster environments as well and will be addressed in Section \ref{sec:disaster}.}.

In practice, the soil medium is typically heterogeneous due to the variations of its content, e.g. amount of sand and water. Hence, the signal propagation may vary as well. In many cases, the ground medium is layered, i.e. neighboring areas along a certain direction or a plane show similar content and therefore similar signal propagation characteristics \cite{wait1951magnetic}. This property can be exploited for the design of RIS-assisted underground networks, if the signal is focused in the most suitable layer for the communication. However, since the distribution of the soil layers and their individual properties are unknown in general, this task is very challenging and should be tackled via adaptive beamforming.

\section{Industry 4.0}
\label{sec:industrial}
\subsection{Overview}
\begin{figure*}
    \centering
    \includegraphics[width=0.9\textwidth]{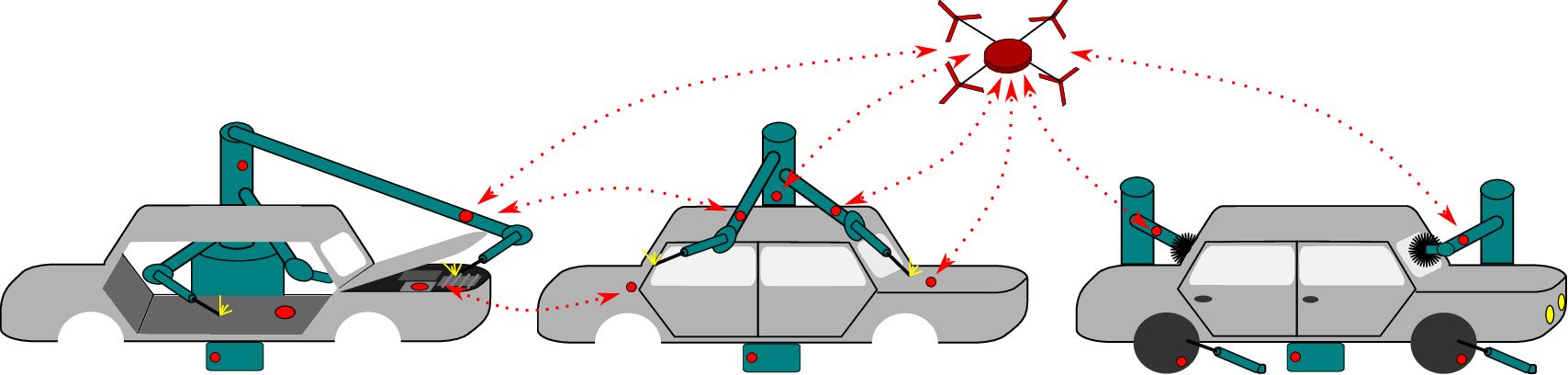}
    \caption{Production line in automobile industry. Sensors monitor the automated process. The data is collected by a UAV.}
    \label{fig:industrial}
\end{figure*}

Industrial applications of wireless communications have gained an increased attention by the research community in the context of the initiative named Industry 4.0. Similarly to the IoT, the concept of Industry 4.0 relies on massive connectivity provided by densely deployed small sensor devices, which are installed in factories and production lines, see Fig. \ref{fig:industrial}.
These sensors are supposed to monitor the accuracy of the manufacturing process and possibly influence this process by providing their feedback to the central processing unit. Through this, the operation of large factories can become fully automated, which may reduce the manufacture and maintenance costs, cf. \cite{7883994}.  

Unlike the previously discussed CEs, the propagation medium in the industrial environment has typically almost no influence on the system performance. Instead, the main challenge for the signal propagation in industrial wireless networks is related to a strong wideband interference generated by the production processes. In addition, a large number of metallic objects as parts of the factory equipment may disturb the transmission. For the massive connectivity of the Industry 4.0, the resulting signal scattering poses a burden, since signal transmissions are impaired by time-variant channels as well as fading effects and multipath. This effect is especially crucial due to the low signal quality from the aforementioned interference. 

Further challenge for the design of the industrial wireless networks is related to very strict requirements regarding point-to-point latency and reliability, since potential malfunction of an industrial process needs to be detected and corrected as fast as possible. Otherwise, the efficiency of production may substantially decrease, cf. \cite{li2017industrial}.


Industry 4.0 and related advances for the wireless communications is an active research area. Unlike the previous use cases, i.e. IoUwT and IoUgT, there is a clear demand and interest of the industrial stakeholders and their involvement in the continuous evolution of this research field. More information about it can be found in \cite{li2017industrial,LU20171,7883994}. 

\subsection{RIS in Industrial Environments}
Industrial environment is best characterized by a large number of obstacles as parts of the infrastructure. These obstacles are responsible for a high degree of signal scattering, which leads to a high path loss as well as severe multipath propagation. Furthermore, some of the target receivers may be blocked from the signal reception and have only a very little field of view for the communication. In addition, the interfering signals from the production lines need to be suppressed. These challenges need to be tackled using RISs. Note that unlike with traditional active relays, no temporal processing of the reflected signal is done by RIS, such that the typical ultra-low latency restrictions of the industrial IoT are not violated. 

We focus on traditional EM signaling in the moderate and high frequency ranges, i.e. mmWave and THz. Especially the high frequency bands of the mmWave seem very attractive for the use in industrial environments \cite{pielli2018potential}. Alternative signaling techniques, i.e. MI or acoustic seem to be much less efficient due to shorter transmission distances and interference, respectively.
\subsubsection{General approach}
\begin{figure*}
    \centering
    \includegraphics[width=0.9\textwidth]{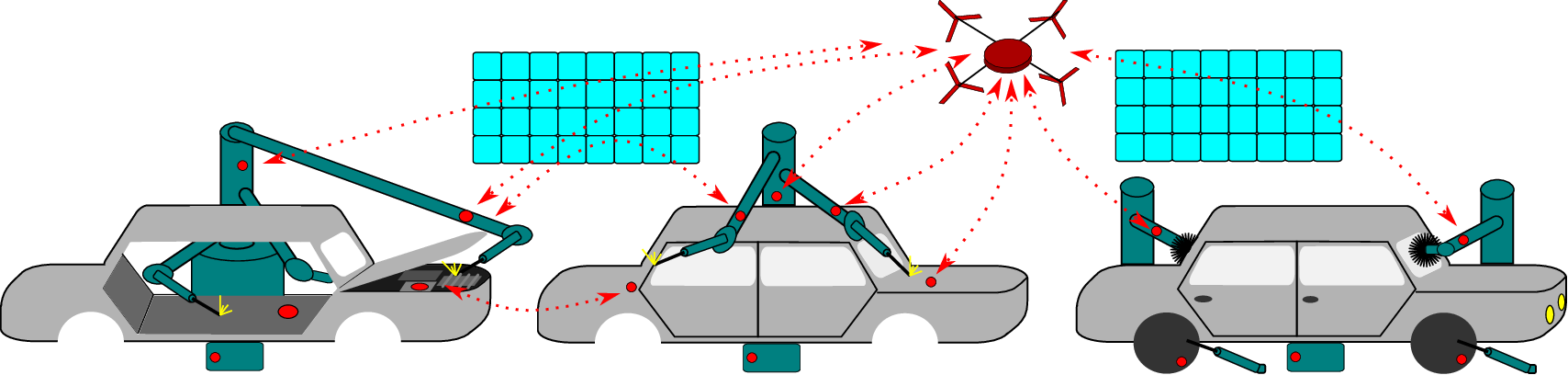}
    \caption{Production line control assisted by RIS.}
    \label{fig:industrial_RIS}
\end{figure*}

The application of RIS in industrial environment is illustrated in Fig. 
\ref{fig:industrial_RIS}. 
Apparently, this scenario has some similarities with the underground communications and specifically with the communication in tunnels and mines, where the scattering is as severe as in the industrial environment. Hence, a similar approach for the signal steering can be selected in this case. The reflective elements of the RISs need to be optimized in order to steer the signal on the shortest path around the obstacles towards the target node. However, in case of industrial environment, the nodes may be deployed in inaccessible areas, such that it might be extremely difficult to find the right phase shift in order to properly navigate the signal. This problem is even more challenging, if discrete phase shifts are used, which substantially reduces the degree of freedom for the optimization.

Additionally, strong interference from some of the production processes should be reduced by destructively aligning the interference signal paths at the receiver. This is especially challenging, since no channel estimation can be performed for the interfering signals, such that only iterative adaptation methods can be designed.

Furthermore, some parts of the infrastructure can be mobile according to the respective automated process. The implications of such moving objects are discussed in the following.
\subsubsection{Moving industrial infrastructure}
Industrial processes usually have a property of repetitive execution of the same actions in order to produce multiple copies of the same product. This can be exploited for the design of RIS-assisted communication networks even in presence of mobile infrastructure. In particular, the repetition of the infrastructural motion manifests as correlated (in time) and deterministic communication channels due to the repetitive positions of the obstacles. The same holds for the positions of mobile RISs, if they are attached to the mobile infrastructure.

This property is very relevant for the design of RIS-assisted networks in this environment, since it substantially eases the synchronization and control of RIS. Also, the requirements for the optimization and update of the phase shifts can be relaxed, which is very beneficial with respect to the ultra-low latency requirements of industrial IoT. In order to exploit the predictable channel characteristics, the statistical features of the received signals can be described using Hidden Markov Models (HMMs), such that the well-known methods of signal detection and synchronization for HMMs can be directly employed here. Alternatively, communication channels can be learned using the well-known methods of artificial intelligence and machine learning. Furthermore, the scheduling of transmissions can be adapted to the repetitions of the channel, which pertain to a high signal quality. Through this, the performance compared to other scheduling methods would significantly improve. 

\section{Emergency networks in disaster areas}
\label{sec:disaster}
\subsection{Overview}
\begin{figure}
    \centering
    \includegraphics[width=0.48\textwidth]{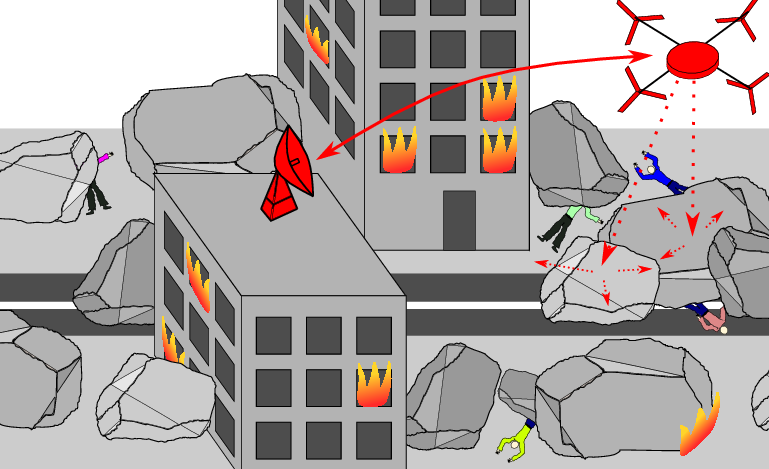}
    \caption{Collapsed buildings and humans trapped under rocks. UAV assists rescue operations. Signals mostly scattered at the broken infrastructure reducing the chances of localizing the humans.}
    \label{fig:emergency}
\end{figure}

Disasters related to earth quake, tsunami, landslide, etc. bring about the destruction of the human habitat. Specifically, the infrastructure of whole cities can be destroyed by such natural causes. On the other hand, human activities like gas explosions can lead to similar results leaving behind collapsed buildings and human bodies trapped inside, see Fig. \ref{fig:emergency}.

The search and rescue (SAR) of humans is the first and most important task to be performed after the disaster. In this context, the destroyed infrastructure is one of the main challenges for the SAR, since it is difficult to search for people underneath. Furthermore, the coordination of the rescue teams and potential application of steerable rescue tools like robots, UAVs, etc. depend on connectivity between the network participants and on reliable signal transmission, cf. \cite{kantor2003distributed,erdelj2016uav,waharte2010supporting}. However, the communication infrastructure is likely to be destroyed in the disaster event as well, which substantially impacts the connectivity and operability of the dedicated wireless network. 

Another challenge for the operation of the wireless network (even if the infrastructure is operational) is the large amount of obstacles, i.e. pieces of the destroyed homes, trees, etc. These obstacles provide a high degree of scattering, such that the resulting interference among the adjacent network links can render the communication unfeasible and unreliable.

Various types of emergency networks to be used in case of disaster have been proposed in the recent years. However, many of the proposed solutions assume that parts of the infrastructure remain whole and capable of assisting the rescue operations. Even though such assumptions are not always valid, the challenge related to the signal scattering is very difficult to assess without these assumptions. 

With increasing popularity of mobile robots, the rescue missions become more and more advanced. For more information on the design of emergency networks in disaster areas, we refer to \cite{JAHIR20191,7057128}.

\subsection{RIS in Disaster Environments}
As mentioned earlier, wireless communication in disaster environments suffers from the scattering and reflections from broken infrastructure, i.e. pieces of buildings, trees, etc. Note that the disaster can occur in any medium including underwater and underground. However, the impact of the disaster on the existing infrastructure is higher on the earth surface, since dense medium can partially absorb the energy from the disaster and reduce the force applied to the infrastructure. Correspondingly, the infrastructure deployed in dense media is typically less deformed by possible disasters or may even partially remain unchanged. 

Hence, we focus on disasters on the ground surface, where the traditional EM signaling in the moderate frequency range is the most promising technique, since it can provide connectivity over large distances (unlike acoustic and MI signaling) and penetrate many types of obstacles (unlike optical signaling).
The challenges here are:
\begin{itemize}
    \item rich scattering of signals,
    \item use of partially broken infrastructure.
\end{itemize}
Both challenges should be resolved using the specifically designed RIS devices.
\subsubsection{General approach}
In order to improve the performance of the mentioned wireless networks supporting the rescue operations in disaster environments, RIS can be deployed as part of the original infrastructure, where the future rescue operations would potentially take place. An example of this infrastructure is the mentioned concept of Smart Cities, where most of the buildings are covered by the reflective surfaces in order to facilitate the massive wireless connectivity. This approach is very beneficial also with respect to the potential danger of a disaster, since parts of the surfaces may still be operational and thus help improving the signal quality of the wireless network established specifically for the SAR operations. 

During the disaster, the infrastructure, e.g. smart buildings, may break in pieces. This includes also the RIS as part of this infrastructure. After the disaster, the separated parts of RISs are no longer functional with existing RIS technology. Nevertheless, it is possible to specifically design RIS in order to account for such scenarios, where parts of RIS need to be easily detached from the main surface.
\begin{figure}
\centering
\begin{subfigure}[t]{0.2\textwidth}
\includegraphics[width=\textwidth]{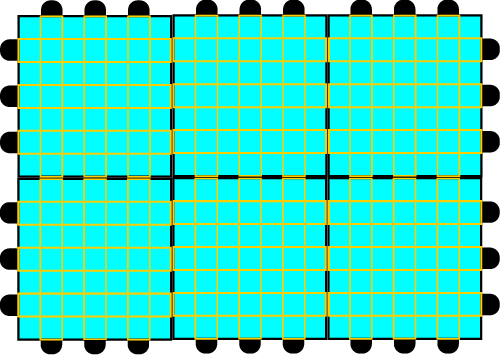}
\caption{Co-located operation.}
\end{subfigure}
~
\begin{subfigure}[t]{0.25\textwidth}
\includegraphics[width=\textwidth]{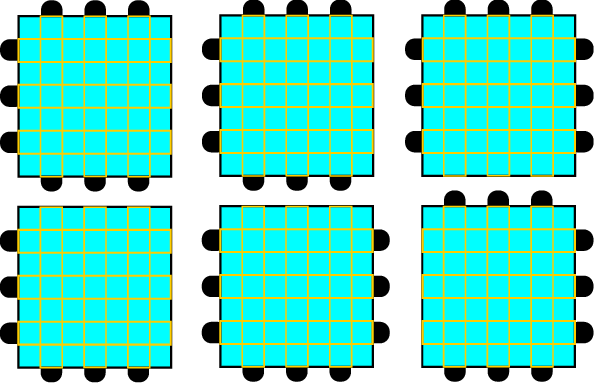}
\caption{Separated operation.}
\end{subfigure}
\caption{Separable RIS: parts of the RIS can easily detach in order to avoid damage and preserve the functionality of the individual parts.}
\label{fig:modes}
\end{figure}

Such separated RIS parts are expected to have sufficient intelligence to provide a basic RIS functionality for a small set of corresponding RIS elements. During the main RIS operation, the parts of RIS are combined to one surface. On the other hand, the parts should be easily detachable in a controlled manner in order to not damage the main processors or the steering mechanisms. One possibility to achieve this is using magnetic connectors. The main benefit of these connectors is the ease of deployment and their robustness.

\begin{figure}
    \centering
    \includegraphics[width=0.48\textwidth]{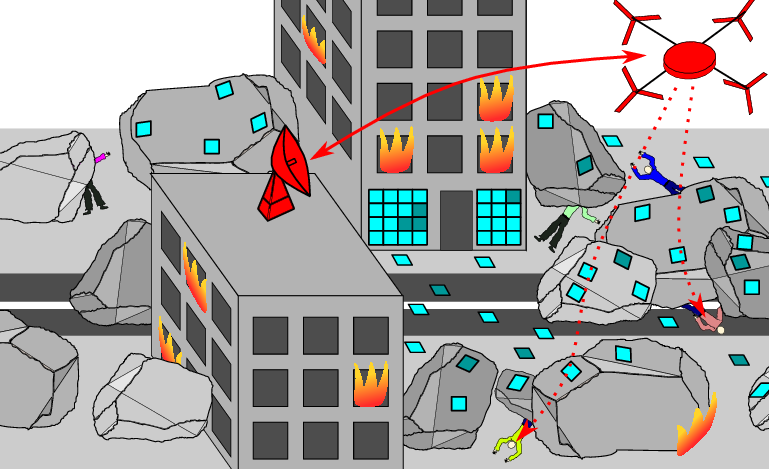}
    \caption{Detached RIS patches support the data collection during rescue operations. Dark colored patches indicate malfunction. Cooperative operation of multiple RIS patches improves the connectivity and the localization.}
    \label{fig:emergency_RIS}
\end{figure}

Specifically, there are two modes of RIS operation, which are separated in time by the disaster event: co-located operation and separated operation. These modes are illustrated in Fig. \ref{fig:modes}. The application of this strategy for the rescue operations is illustrated in Fig. \ref{fig:emergency_RIS}. 
In the following, we discuss the potential research challenges for each mode.
\subsubsection{Co-located operation}
We consider a large RIS, which consists of multiple smaller RIS patches connected to each other. The large RIS should be able to exploit the processing capabilities, memory and stored energy of all its parts to provide the best possible performance. 

The joint processing can be performed in a distributed manner, i.e. similar to edge computing. In order to maintain the connectivity between individual RIS patches, one of the patches may take the role of the master node of the local patch cluster. This master node would manage the operation of all connected slave patches. The management of the respective master-slave connections can be achieved using one of the existing protocols, e.g. on the basis of Bluetooth or ZigBee technology.

For the RIS optimization, the objective in this mode is mostly related to the beam steering in order to maximize the receive signal power or enhance the secrecy.
\subsubsection{Separated operation}
After the separation, the distances between the patches of RIS may be substantially larger, such that it probably does not make sense to consider them as parts of the same entity. Instead, each RIS patch can be assumed to have only very limited number of reflective elements and shall be responsible only for its own operation. On the other hand, some RIS patches may find themselves in a close proximity to others, such that distributed computation may still be possible. Correspondingly, these patches would preferably reconfigure and build a new edge computing cluster. In order to determine, which RISs should form the cluster, each patch should possess the sensing capabilities or be able to localize itself. This methodology is also applicable in underground environment, as mentioned earlier, where the cost of deploying large RISs may be very high. Instead, clusters of small RISs can be employed.

In this mode, i.e. after the disaster event, the objective for the phase shift optimization of each individual RIS patch is related to the reduction of the scattering effect and the harmful multipath propagation. Hence, the individual RIS patches should act as ad hoc nodes and collaborate with each other in order to solve this task cooperatively. A similar challenge for the ad hoc networking of RIS relays has already been identified in \cite{kisseleff2020reconfigurable}. In case of disaster environment, the difficulty is even higher due to the reduced memory, energy and processing capabilities of the small RIS patches.

\section{System design aspects}
\label{sec:system}
In this section, we propose a basic hardware architecture applicable in the target scenarios and a generic channel model for RIS-assisted signal transmission for wireless networks in CEs. The latter is numerically evaluated for two selected scenarios of considered CEs.
\subsection{Hardware architecture}
\label{sec:hardware}
As recognized in various works, e.g. \cite{kisseleff2020reconfigurable}, future RIS-assisted wireless networks will heavily depend on the autonomous operation of RIS. While this will remain an optional feature for the traditional wireless networks, it will be certainly a necessary condition for the operation of RIS in CEs, since most of the RIS devices will be hardly accessible after the initial deployment. Hence, each RIS may not only have a passive functionality of signal reflection, but also an advanced functionality of a wireless sensor/actuator. Hence, future hardware design will substantially differ from the ones discussed in the references of Section \ref{sec:2_new}. In the following, we propose a novel architecture, which should enable this advanced functionality.

\begin{figure}
    \centering
    \includegraphics[width=0.48\textwidth]{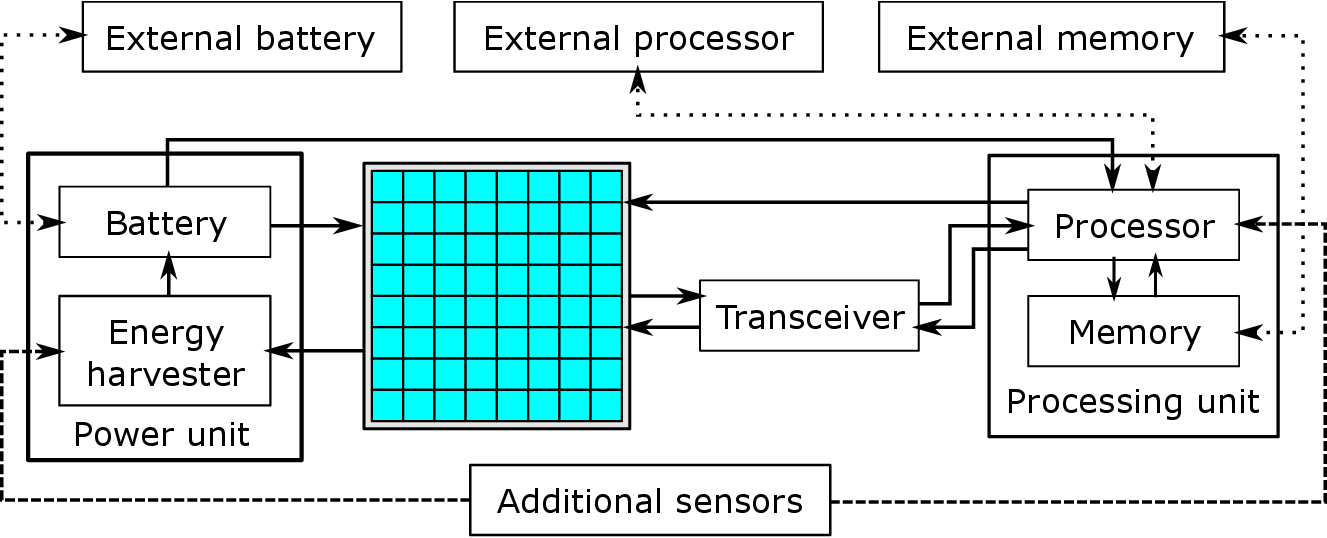}
    \caption{Proposed generic architecture for RIS devices in CEs.}
    \label{fig:hardware}
\end{figure}

Based on the traditional architecture of wireless sensor nodes, cf. \cite{akyildiz2006wireless}, it is possible to obtain a generic architecture of a RIS node for the application in CEs, see Fig. \ref{fig:hardware}. In addition to RIS employed as a front-end, each node may be also equipped with traditional antennas depending on the system requirements and scenario. In particular, such additional antennas can be employed in order to enable more advanced functionalities, such as active relaying, signal buffering or frequency conversion, without changing the behavior of the reflective elements of RIS. In this case, RIS and the antennas are operated in parallel and have identical connections to other hardware components. Similarly, RIS can be connected to additional sensors in order to be able to sense the environment and perform signal routing by adjusting the impedances.

The power unit contains a battery which is used to power all other units and to manipulate/activate the reflective elements of RIS. Since energy resources are very scarce in CEs, energy harvesting modules can be utilized to harvest the signals, which are not reflected (i.e. absorbed) by RIS. This strategy is well aligned with the recently proposed term of 'self-sustainable' RIS \cite{9214497}. Typically, only parts of the signals are utilized for energy harvesting in order to maintain the connectivity of the network. Hence, SWIPT technology seems promising, which may utilize power splitters prior to energy harvesting. Alternatively, one cluster of reflective elements can be assigned to signal reflection (or reception of the control signaling) and another cluster to energy harvesting. Furthermore, in order to account for the proposed co-located operation of multiple RIS patches, each power unit may have connections to external batteries, e.g. batteries of the adjacent RIS patches or other energy sources.

The phase shifts are set by the processing unit, which consists of a processor and a memory storage. In addition, RIS can be used as a transmitter according to \cite{9133266} via a specifically designed transceiver connected to the processor. Similarly, the receiver functionality can be enabled in order to receive the control signals and configuration parameters from other RISs. In order to enable more advanced signal processing capabilities, processing unit can be connected to external processors and memory blocks of adjacent RIS patches. A distinct advantage of this strategy is the reduced computational complexity per RIS and the promising feature of functional extendability of each RIS.

\subsection{Signal propagation}
\label{sec:channel}
For the basic system model, we consider a single point-to-point (P2P) connection using single antenna transceivers in a CE with multiple RISs. Note that this is different from most of the works on RIS, where the spatial diversity is enhanced by utilizing multiple antennas at the transmitter or at the receiver. As mentioned earlier, the main challenge in CEs is not to exploit the spatial diversity to boost the achievable data rate, but to improve the connectivity of a wireless network despite very harsh propagation conditions. Furthermore, the employed transceivers can belong to small wireless sensor nodes, which can only be equipped with a single antenna due to complexity and space limitations.

In addition, we focus on frequency-selective channels, which are typical for CEs. In this case, the signal propagation from one transceiver to another can be described using the following classical equation:
\begin{equation}
    y(t)=(h_{d}(t)+h_{\boldsymbol{\theta}}(t))*x(t)+w(t),
\end{equation}
where $x(t)$ and $y(t)$ represent the transmitted and received signals, respectively. The variance of $x(t)$ corresponding to the transmit power is denoted as $\sigma_x^2$. Here, '*' denotes the convolution operation. In addition, $w(t)$ represents a realization of an additive noise with the variance $\sigma_{w}^2$, which depends on the type of employed signaling. In addition, the direct link between the transmitter and the receiver in CEs may not be negligible. However, due to the multipath propagation in CEs, this direct link does not necessarily correspond to a line-of-sight propagation. It is thus described via channel impulse response $h_d(t)$, which comprises a set of multipath components. Since the receiver locks on the dominant channel tap (both for the direct link and for RIS-aided transmissions), all other channel taps pertain to the interfering signals, which reduce the signal quality. Accordingly, we denote $\bar{h}_d$ and $\tilde{h}_{d, i},\forall i$ the channel gain associated with the dominant signal path and the channel associated with the $i$th multipath component, respectively, of the direct link. 

The channel impulse response pertaining to the signals reflected from the RIS is denoted by $h_{\boldsymbol{\theta}}(t)$ and is a function of the parameters $\boldsymbol{\theta}_{n}\in\mathbb{C}^{Q_{n}\times 1}$, with $n\in\{1, \dots, N\}$, comprising the reflection coefficients associated with the $Q_{n}$ reflective elements/planes of the $n$-th RIS. It is assumed that the $q$-th complex reflection coefficient of the $n$-th RIS can be written as
\begin{align}
    \theta_{n, q} \!= \eta_{n, q} \!\cdot{\rm e}^{{\rm j}\phi_{n, q}}\!,\forall (n,q)\in\{1,\dots,N\}\times\{1,\dots,Q_{n}\},
\end{align}
with $\phi_{n,q} \in [0,2\pi)$ denoting the corresponding phase shift, and $\eta_{n,q} \in [0,1]$ denoting the corresponding reflection efficiency. 

In case of multiple RISs deployed in the same environment, it is important to consider multi-hop reflections as well.  Specifically, some signals arriving at the $n$-th RIS can be reflected towards $k$-th RIS, which either shortens the overall signal path or prolongs it. In the latter case, the signals will be substantially more delayed and more attenuated due to additional path losses. Hence, such multipath components are likely to be negligible.

\subsection{Signal quality improvement}
While the main challenge for the network operation in CEs remains the connectivity, as mentioned earlier, the signal quality can be used to demonstrate the potential improvements in signal propagation. In fact, the connectivity can be addressed using various metrics, such as outage probability, communication range of each node (i.e. maximum distance, for which a certain signal quality can be guaranteed), network throughput, etc. \cite{akyildiz2010wireless}. All these metrics are directly related to the performance of individual network links in terms of achievable rate or signal quality. In the following, we focus on signal-to-interference-plus-noise ratio (SINR) as a key performance indicator.

We consider a single RIS in our investigation. A two-hop RIS-assisted wireless link can be thus described via channel impulse response given by
\begin{eqnarray}
    \label{eq:channel}
    h_{\boldsymbol{\theta}}(t)=\sum_{q=1}^{Q_1}\left(h^{\mathrm{Tx->RIS}}_{1,q}(t)*h^{\mathrm{RIS->RX}}_{1,q}(t)\right)\theta_{1,q},
\end{eqnarray}
where $h^{\mathrm{Tx->RIS}}_{1,q}(t)$ and $h^{\mathrm{RIS->RX}}_{1,q}(t)$ represent the sub-channels connecting transmitter with RIS and RIS with receiver, respectively.
Each of the sub-channels may comprise a different set of multipath components and can be associated with a different frequency selectivity. As an example, some experimental measurements obtained for underwater acoustic channels suggest that the multipath components follow a Rician, Rayleigh or $K$-distribution, cf. \cite{stojanovic2009underwater}. For the target applications, we can assume Rician fading with a large variance of the random components. Such fading models are also partially applicable in other scenarios of CEs. If each multipath component is treated as interference, it is possible to express the channel impulse response as $h_{\boldsymbol{\theta}}(t)=\bar{h}_{\boldsymbol{\theta}}(\tau_0)+\tilde{h}_{\boldsymbol{\theta}}(t)$, where
\begin{equation}
\label{channel_deterministic}
\bar{h}_{\boldsymbol{\theta}}(\tau_0)=\delta(t-\tau_{0})\sum_{q=1}^{Q_1}\left(h^{\mathrm{Tx->RIS},0}_{1,q}h^{\mathrm{RIS->RX},0}_{1,q}\right)\theta_{1,q}
\end{equation}
is the deterministic part associated with the main tap of the channel $h_{\boldsymbol{\theta}}(t)$ and
\begin{equation}
\label{channel_stochastic}
    \tilde{h}_{\boldsymbol{\theta}}(t)=\sum_{q=1}^{Q_1}\sum_{m=1}^{M_{1,q}}\tilde{h}_{1,q,m}\delta(t-\tau_{q,m})
\end{equation}
is the stochastic part associated with the interfering multipath components, where $\tilde{h}_{1,q,m}\in\mathbb{C}$ denotes the $m$th multipath component of the signal reflected from the $q$th RIS element. Here, $\delta(t-\tau_{0})$ corresponds to the delay of the main channel tap by $\tau_{0}$ and $\delta(\cdot)$ denotes the Dirac pulse. Accordingly, $\delta(t-\tau_{q,m})$ represents the time shift relative to the dominant tap by $\tau_{q,m}$. Note that $\left|h^{\mathrm{Tx->RIS},0}_{1,q}h^{\mathrm{RIS->RX},0}_{1,q}\right|\approx g,\forall q$ with a suitable $g$, since the path length is almost equal for all signal paths independently of the reflective element. Let $\sigma_{\tilde{h}}^2$ denote the variance of the stochastic channel components $\tilde{h}_{1,q,m},\forall q,m$. Due to the circular symmetry of these channel taps the received signals from all these taps can be approximately viewed as uncorrelated and the corresponding mean value is zero in this case. Furthermore, these channel taps are not correlated with the random variable $x(t)$, such that the variance of each received signal component $\tilde{h}_{1,q,m}x(t)$ is given by $\sigma_{\tilde{h}}^2\sigma_x^2$ according to \cite{frishman1975arithmetic}. Thus, we obtain the expected interference power by summing up the variances and adding them to the interference power of the direct link
\begin{eqnarray}
    I\hspace*{-2mm}&=&\hspace*{-2mm}\mathcal{E}\left\{\left|\tilde{h}_{\boldsymbol{\theta}}(t)*x(t)+\sum_i\delta(t-\tau_{i})\tilde{h}_{d, i}*x(t)\right|^2\right\}\notag\\ \hspace*{-2mm}&=&\hspace*{-2mm}\sigma_{\tilde{h}}^2Q_1M_{1,q}\sigma_x^2+\sum_i|\tilde{h}_{d, i}|^2\sigma_x^2,
\end{eqnarray}
where $\mathcal{E}\{\cdot\}$ denotes the expectation operation.
Note that the consecutive transmitted symbols are statistically independent, such that the sum of a large number of weighted adjacent symbols can be treated as an uncorrelated disturbance with signal characteristics close to that of an AWGN according to the Central Limit Theorem.

In contrast, if the phase shifts $\theta_{1,q}$ are optimized to constructively overlap at the dominant tap, we obtain for the channel power gain of this tap
\begin{equation}
    S=\mathcal{E}\left\{\left|\left(\bar{h}_d+\bar{h}_{\boldsymbol{\theta}}(\tau_0)\right)x(t)\right|^2\right\}\approx (gQ_1+\bar{h}_d)^2\sigma_x^2.
\end{equation}
Hence, the SINR is obtained via
\begin{equation}
    \mathrm{SINR}=\frac{S}{I+\sigma_w^2}
    \approx \frac{(gQ_1+\bar{h}_d)^2\sigma_x^2}{\sigma_{\tilde{h}}^2Q_1M_{1,q}\sigma_x^2+\sum_i|\tilde{h}_{d, i}|^2\sigma_x^2+\sigma_{w}^2}.
\end{equation}
From this equation, we can deduce that for a large number of RIS elements $Q_1$ the SINR can be improved by up to 10 dB per decade. On the other hand, depending on the signal strength of the direct link, we may obtain a signal quality improvement of up to 20 dB per decade, if $\sigma_{\tilde{h}}^2Q_1M_{1,q}\sigma_x^2\gg\sum_i|\tilde{h}_{d, i}|^2\sigma_x^2+\sigma_{w}^2$ holds. This makes the use of RIS in such scenarios especially promising.

\subsection{Numerical evaluations}
In this section, we evaluate the performance of RIS-assisted communication in all four environments. 

\begin{figure}
    \centering
    \includegraphics[width=0.48\textwidth]{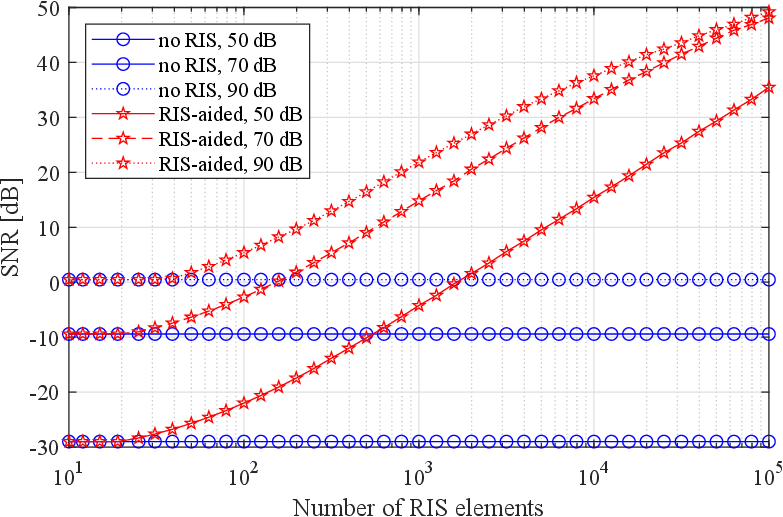}
    \caption{SINR vs. number of RIS elements for different transmit SNR values $\sigma_x^2/\sigma_w^2$ in shallow underwater scenario.}
    \label{fig:underwater_SNR}
\end{figure}

We start with an acoustic underwater system based on the channel model from \cite{stojanovic2009underwater}, where the transmitter and the receiver are deployed close to the ground at 50 m depth and are 1000 m apart. We assume a spreading loss of 1.5 and a carrier frequency of 100 kHz. Also, 3 dB loss is added to each reflection from the ground while reflections from the RIS and the surface of the sea are considered free of losses. Furthermore, the ratio $\sigma_x^2/\sigma_w^2$ corresponding to the assumed transmit SNR is used as a parameter. Here, we assume that each RIS element creates an additional path between the transmitter and the receiver. The reflections at RIS are optimized to amplify the strongest path via passive beamforming. The results are depicted in Fig. \ref{fig:underwater_SNR}.
We observe that the performance can be dramatically improved using RIS with a large number of elements. Also, with increasing transmit SNR, the SINR increases both with and without RIS. With large SNR and high number of RIS elements, the inclination is 10 dB per decade, whereas with low SNR or low number of RIS elements, it is 20 dB per decade.

\begin{figure}
    \centering
    \includegraphics[width=0.48\textwidth]{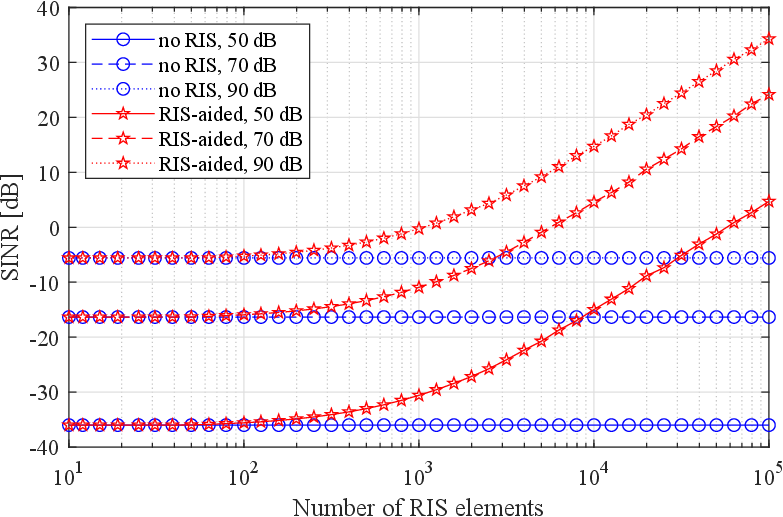}
    \caption{SINR vs. number of RIS elements for different transmit SNR values $\sigma_x^2/\sigma_w^2$ in underground tunnel scenario.}
    \label{fig:underground_SNR}
\end{figure}

Next, we investigate an RF based underground communication system deployed in a tunnel of 10 m width. Both transmitter and receiver are assumed to be mounted in the same wall of the tunnel 100 m apart. Here, we select the path loss exponent 4.3 and focus on reflections according to the tunnel geometry. The results are depicted in Fig. \ref{fig:underground_SNR}.
Similarly to the underwater scenario, we observe an improvement of the SINR with increasing number of reflective RIS elements and with the transmit SNR. However, only the increase by 20 dB per decade is observed in this scenario, which is very beneficial.

\begin{figure}
    \centering
    \includegraphics[width=0.48\textwidth]{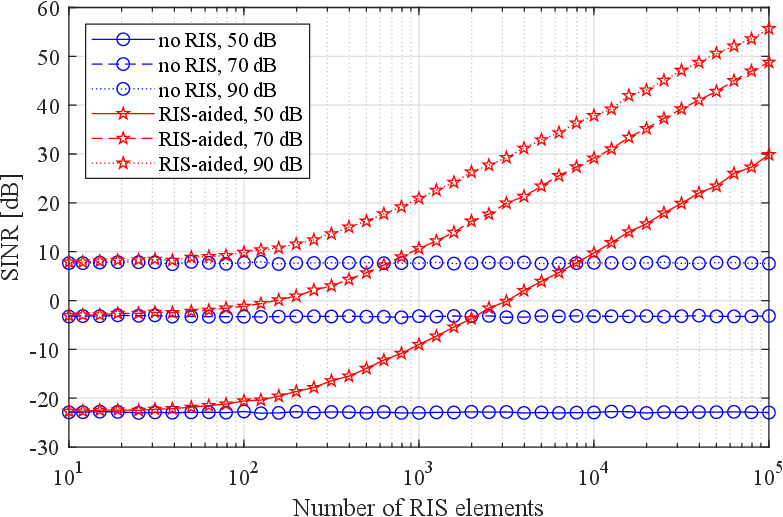}
    \caption{SINR vs. number of RIS elements for different transmit SNR values $\sigma_x^2/\sigma_w^2$ in industrial scenario.}
    \label{fig:industrial_SNR}
\end{figure}

For the industrial environment with rich scattering, we may observe many time-shifted Rician distributed (with Rician factor 0 dB) multipath signal components. The choice of the Rician fading is due to the likely optimized placement of the network nodes in controlled industrial environment, such that the probability of a line-of-sight path is very high. In addition, we can assume that due to the rich scattering the distorted signal repetitions would arrive in each symbol interval. However, the path loss corresponding to each of these repetitions pertains to the transmission distance that the signal traveled during the time between the departure and the arrival, such that the path loss gradually increases over time, which also limits the number of relevant signal repetitions. Obviously, this type of channel differs from the two channels described above. The distance between the transmitter and the receiver is set to 50 m, the path loss exponent is 4.3 and the sampling frequency is 10 MHz. The RIS is placed in the middle between the two nodes and its elements are optimized to maximize the signal strength of the strongest channel tap. The results are depicted in Fig. \ref{fig:industrial_SNR}.
Again, we observe an improvement of the SINR with increasing number of reflective RIS elements and with the transmit SNR. For a large number of RIS elements and low SNR, the SINR improvement is 20 dB per decade. However, with SNR = 90 dB and starting from $10^4$ elements, the gain slightly decreases to 18 dB per decade, which is similar to the underwater scenario.

\begin{figure}
    \centering
    \includegraphics[width=0.48\textwidth]{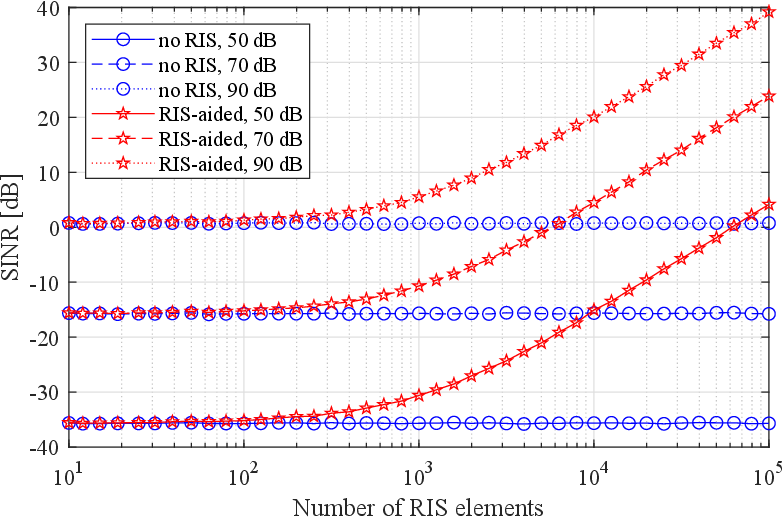}
    \caption{SINR vs. number of RIS elements for different transmit SNR values $\sigma_x^2/\sigma_w^2$ in disaster scenario.}
    \label{fig:disaster_SNR}
\end{figure}

Consider now a disaster environment, which is characterized by many time-shifted multipath components. Here, we assume a Rayleigh fading, since the probability of blockage for any communication link is very high. Similarly to the industrial environment, we assume that the signal repetitions would arrive in each symbol interval. The distance between the transmitter and the receiver is set to 100 m, the path loss exponent is 4.3 and the sampling frequency is 10 MHz. The RIS is placed in the middle between the two nodes and we assume a co-located mode of RIS operation for simplicity. The resulting signal quality is depicted in Fig. \ref{fig:disaster_SNR}.
Here, the same observations can be made as in industrial and underground scenarios, i.e. the signal quality improves with increasing number of reflective RIS elements with 20 dB per decade, which leads to very high gains.

Correspondingly, it is shown that the deployment of RIS in considered challenging environments can be very beneficial.

\section{Future challenges}
\label{sec:4}
The research challenges can be split in two categories:
\begin{itemize}
    \item general RIS-related challenges,
    \item specific RIS-related challenges associated with the deployment in the CEs.
\end{itemize}
At first, we briefly review the general challenges associated with RIS and then  we discuss the specific challenges for the respective use cases specified in the previous section.
\subsection{Known challenges for RIS}
\label{sec:3}
Like in the terrestrial RIS-assisted communication networks, the main challenges can be outlined as
\begin{itemize}
    \item channel estimation/pilot decontamination,
    \item precoding/beamforming,
    \item control signaling.
\end{itemize}
Channel estimation is one of the biggest challenges for RIS-assisted wireless networks. 
This problem becomes even tougher with increasing number of RIS elements and with increasing number of RIS devices, for which the estimation needs to be performed. In addition, similarly to the Massive Multiple-Input Multiple-Ouput (MIMO) systems, the channel estimation is impacted by the correlation between training sequences in terms of pilot contamination. Various types of channel estimation for the whole cascade of RIS links have been analyzed so far, see Section \ref{sec:sota}. However, regarding the time-variant and higly dispersive channels in CEs, these methods may not be sufficiently accurate.

Precoding and beamforming are considered the key functionality of RIS. Despite a large number of research works dedicated to this system aspect, the design rules for the precoding/beamforming in wireless networks with multiple RISs is not yet fully understood. In fact, passive beamforming using multiple RISs with ping-pong multihop signaling seems very challenging. Interestingly, the channel correlation in CEs is very rare due to the various effects impacting each individual communication link, such that the spatial diversity can be better exploited. However, to combat the path loss, large RISs with many elements are needed, which substantially enhances the complexity of the phase shift optimization.

Control signaling is important for the operability of RIS. In order to update the phase shifts of the RIS elements, the connectivity between RIS and its parental node (e.g. base station) has to be very reliable and fast. This becomes a burden in case of a large number of elements, since the corresponding large amount of data may require a very high data rate to ensure that the update frequency satisfies the system requirements. For the CEs, the control signals will be highly distorted and attenuated, which may cause errors in their decoding and potential need for retransmission.

Although similar concepts as already proposed in the literature (and mentioned in Section \ref{sec:sota}) for the respective terrestrial networks can be applied in the CEs, these methods may need to be adapted to the time-variant and dispersive channels as well as a typically much lower signal strength, which leads to an overall extremely low signal quality.

Further traditional challenges for the design and optimization of RIS technology in future wireless networks have been well described in the previous works and surveys. Hence, we refer to \cite{Gong,di2020smart,kisseleff2020reconfigurable} for more information.
In addition to the specific challenges identified for each use case in the previous section, we consider the most promising future research challenges:
\begin{itemize}
    \item deployment planning and automation,
    \item distributed RIS operation,
    \item powering of RIS.
\end{itemize}
These open problems are discussed in the following.
\subsection{Deployment planning and automation}
The strategic RIS deployment has been considered in some works so far in the context of future wireless networks. Furthermore, the RIS-equipped UAVs have been introduced in order to find the optimal position for the RIS. This optimal position may depend on the scenario as well as on the time-varying channel conditions and available power resources.

In the context of CEs, the strategic deployment of RISs is of paramount importance for the effectiveness of RIS due to the high complexity and spatial selectivity of the communication channels.

Specifically, the transmission distances in underwater environment are rather large, which leads to a large path loss both for the main signal tap and for the signal reflections. Correspondingly, by placing RISs in certain areas, the signals reflected from the RIS may be impaired by a higher pathloss than the signals reflected from other obstacles. In this case, RIS would not always be able to compensate the multipath components. 

For the deployment in the tunnels and mines, we expect a large number of ping-pong reflections for the signal redirection. However, each reflection may contribute implies additional path loss due to partial absorption of the signal. Hence, the strategic deployment of RISs is needed in order to reduce the number of reflections. This would lead to a better overall signal quality.

In disaster environments, it is difficult to predict the distribution of RISs e.g. after the explosion. Hence, it is necessary to put RIS in as many locations as possible in order to make sure that the expected density of the RIS parts does not undergo the required minimum.

In general, the path finding optimization for the signal propagation is required in order to identify the best possible locations for the RIS deployment.

CEs are additionally characterized by the difficulties associated with the deployment of the network nodes. The reason is that it might be difficult for the humans to reach the deployment location and perform the deployment work. Hence, this process needs to be automated.

Specifically, the deployment procedure in the underground and underwater media as well as crowded factories will make use of robots to carry RISs and attach them to the dedicated surfaces. In this context, the automation problems related to the design of such robots are of interest.
For the disaster environments, the deployment automation is less important, since it is similar to the traditional deployment strategies for the smart cities.
 
\subsection{Distributed RIS operation}
\label{sec:distributed}
As mentioned in the previous works, distributed operation of RIS can be very beneficial, especially with respect to the discussed application in disaster environments. Distributed RIS operation requires that each RIS or its part should be attached to a sensor node with respective capabilities, i.e. memory, signal processor, battery, etc. Correspondingly, the entire set of wirelessly connected RISs needs to be arranged in wireless sensor/actuator network. Subsequently, the wireless networking aspects need to be investigated, in particular the higher OSI layers, such as network and transport layers. 

In this context, routing and topology control for the RIS network (not to confuse with the target network, for which the smart environment is created) are of special interest. These aspects are difficult to optimize due to a complex dependency between the communication environment and the optimization of the phase shifts for the target application. Hence, it seems that the most promising design method will be based either on metaheuristics or on game theory. The latter seems promising due to the ad hoc behavior of RIS-equipped nodes.

\subsection{Powering of RIS}
One of the main open problems for the distributed operation of RIS is the recharging of its battery. in the context of CEs, this problem is even more challenging, since the channels to be used for wireless powering can be time-variant with a very short coherence time or extremely dispersive with a high path loss. Accordingly, the power transfer efficiency might be extremely low in most of the cases. Furthermore, in some applications like disaster environments, there is no common energy source for the wireless powering of small sensor devices. Hence, the battery recharging can be done only using one of the following alternative approaches:
\begin{itemize}
    \item each RIS may not only reflect the signals, but also partially absorb their energy and use it for the own recharging. A distinct advantage of this strategy is that RIS would drain the EM radiation from the medium and thus make the environment more green;
    \item RIS should be able to generate power from any type of received signaling, e.g. EM and acoustic radiation, vibration, x-ray, etc.
\end{itemize}
The first strategy implies that the signal routing and the optimization of the reflective elements need to be adapted in order to account for the charging of the RIS batteries. The second strategy implies that various types of sensors need to be attached to each RIS in order to be able to harvest energy from the respective sources. While the first strategy is challenging with respect to the online operation, the second strategy is challenging for the initial RIS design. In future applications, energy scavenging might be considered, which would rely on even more intelligent autonomous operation of RISs.
\section{Conclusion}
\label{sec:5}
In this paper, we presented our vision and potential research challenges for the deployment of RIS in CEs. Specifically, the benefits for the signal propagation in large wireless networks, such as IoT, have been highlighted. Regarding the target scenarios, four most challenging use cases have been investigated, i.e. underwater, underground, industrial and disaster environments. These use cases have been unified under the umbrella of a single enabler, i.e. RIS-assisted networking. In this context, the enabling properties of RIS have been discussed. Specifically, for the underwater and underground media, RIS can be used to reduce the impact of the harmful multipath propagation. In industrial environment, RIS can be used to redirect the signals in order to avoid absorption and reflection by metallic objects. In disaster environment, future RIS can be employed to preserve connectivity and thus assist the SAR operations even after a possible infrastructure damage. 

Potential deployment strategies, hardware design, expected analytical improvements for the signal propagation supported by numerical results have been provided. Furthermore, open research directions for the near future and for the long-term evolution have been presented. Correspondingly, this paper is expected to contribute to the rapid advancement of the research field both for the CEs and for the RIS technology.
\bibliographystyle{IEEEtran}
\bibliography{IEEEabrv,Literature}
\end{document}